\documentclass[12pt,a4paper]{article}
\usepackage{amssymb,amsmath,graphicx,subcaption,cite}

\usepackage{epstopdf}
\usepackage[utf8]{inputenc}
\usepackage[english]{babel}

\begin{document}

	
\begin{center}
   {\Large\bf Monte Carlo study of the phase diagram of layered XY antiferromagnet}
\end{center}
\vskip 1 cm
\begin{center}
    Muktish Acharyya$^{1,*}$ and Erol Vatansever$^{2}$

   \textit{$^1$Department of Physics, Presidency University,}\\
   \textit{86/1 College Street,
           Kolkata-700073, India} \\
   \textit{$^2$Department of Physics, Dokuz Eylul University}\\
   \textit{Tr-35160 Izmir, Turkey}
\vskip 0.2 cm

   {Email$^1$:muktish.physics@presiuniv.ac.in}\\
   {Email$^2$:erol.vatansever@deu.edu.tr}
\end{center}
\vspace {1.0 cm}
	
\noindent {\bf Abstract:} The three-dimensional XY model is investigated in the presence of a uniform magnetic field applied in the $X$-direction. The nearest neighbour intraplanar interaction is considered ferromagnetic, and the interplanar nearest neighbour interaction is chosen to be antiferromagnetic. Starting from a high-temperature initial random spin configuration, the equilibrium phase of the system at any finite temperature was achieved by cooling the system using the Monte Carlo single spin-flip Metropolis algorithm with a random updating rule. The components of total magnetisation and the sublattice magnetisations were calculated. The variance of the antiferromagnetic order parameter and the susceptibility have been calculated. In a specific range of relative strengths of interactions (antiferromagnetic/ferromagnetic) and the applied magnetic field, the system shows the equilibrium phase transitions at different temperatures. The phase diagrams (in the field-temperature plane) were obtained for different values of the relative interaction strengths. The ordered region bounded by the phase boundary was found to increase as the ratio of relative interaction strength increased. Furthermore, the maximum value of the susceptibility ($\chi^m_{ay}$) was found to increase with the system size ($L$). For $\chi^m_{ay} \sim L^{{{\gamma} \over {\nu}}}$, the exponent $\gamma/\nu$ has been estimated to be 2.10$\pm0.11$ .

\vskip 4cm


\vskip 1cm

\newpage
\noindent{\large\bf I. Introduction:}

\vspace {0.5 cm}

Magnetic field-induced phase transitions have been the focus of numerous theoretical and experimental studies in condensed-matter physics and statistical physics over the last several decades. Metamagnetic systems under the influence of an external magnetic field can show unusual and interesting physical behaviours due to the competition between antiferromagnetic and ferromagnetic interactions. More specifically, the spins in the intralayer part of the system interact ferromagnetically with each other while they interact antiferromagnetically with each other along the interlayer part of the system. In the absence of an external field, a typical Ising metamagnetic system has an antiferromagnetic order. When the strength of the field is increased starting from zero, its phase transition point separating the ordered and disordered phases from each other gets shifted to the lower temperature region. From the theoretical point of view, thermal and magnetic phase transition properties of different kinds of metamagnetic systems have been studied by a wide variety of techniques such as Mean-Field Theory \cite{Moreira1, Moreira2, Liu, Wei, Liang, Gulpinar1, Gulpinar2, Nascimento}, Effective-Field Theory \cite{Zukovic0, Zukovic01, Filho, Geng, Miao}, Monte-Carlo simulation method \cite{Landau, Aora, Hernandez, Pleimling1, Pleimling2, Pleimling3, Acharyya1, Zukovic1, Zukovic2, Weinzenmann, Chou, Mayberry}, and High Temperature Series Expansion method \cite{Harbus1, Harbus2}. For example, in Ref. \cite{Mayberry}, the authors have considered the synthetic metamagnetic films within the framework of Monte Carlo simulations. Based on their simulation results, they found three stable phases, where one phase boundary ends in a critical end point, whereas the other phase boundary shows a tri-critical point at which the transition changes from first to second order. According to the Mean-Field Theory predictions, a metamagnetic system is expected to display a tri-critical point decomposition to a critical and a bi-critical end point for a sufficiently small ratio between intralayer ferromagnetic and interlayer antiferromagnetic couplings. This prediction has been tested by Monte Carlo simulations in Ref. \cite{Zukovic1}, and in contrast to the mean field case, it is reported that there is no evidence of such a decomposition and produces a tri-critical behaviour even for a coupling ratio as small as $R = 0.01$. As an interesting example, it is possible to say that the magnetic behaviour of a typical Ising antiferromagnet under the existence of a magnetic field is different from that found in $^3He$-$^4He$ mixtures in the tri-critical region \cite{Landau}. In another work, \v{Z}ukovic and Idogaki focused on the dilution effects on the multi-critical behaviour observed in the metamagnetic systems by performing Monte Carlo simulations \cite{Zukovic2}. They reported that the region where a first order phase transition appears survived down to at least $p=0.5$, where $p$ controls the density of the magnetic sites.

Furthermore, the metamagnetic systems can present multiple peaks in thermodynamic functions such as specific heat and susceptibility curves as a function of the temperature \cite{Pleimling1, Pleimling2, Acharyya1}. Readers may refer to \cite{Stryjewski} for a detailed review of the metamagnetism observed in many systems. We note that, among the many metamagnets, three of the most studied are $FeBr_2$, $FeI_2$ and $FeCl_2$. In addition to the equilibrium properties, many studies have been devoted to elucidating the non-equilibrium features of the metamagnetic systems driven by a time dependent magnetic field \cite{Keskin1, Deviren1, Temizer, Acharyya2}.

Most of these studies mentioned above are performed in discrete Ising type spin models. It is known that continuous classical spin models like XY are of great interest because they have very rich physical behaviors. The order parameter belonging to the XY model has only two components. For the first time, a new definition of order called topological order is proposed by Kosterlitz and Thouless for two-dimensional planar magnets like XY ferromagnets in which no long-range ferromagnetic order exists \cite{Kosterlitz}. Since then, the magnetic properties of different kinds of systems have been handled with the XY model. For instance, spin-dynamics regarding the structure factor and transport properties of the three-dimensional XY model have been investigated by Monte-Carlo simulations \cite{Krech}. Extensive Monte Carlo simulations of the classical XY model in three dimensions containing random site dilution have been performed by using a hybrid algorithm \cite{Filho2}. Based on the numerical outcomes, it is found that the critical exponents and universal cumulants are independent of the amount of dilution. Recently, the XY vectorial generalisation of the Blume-Emery-Griffiths model has been treated on thin films, and its thermodynamical properties are determined as a function of the film thickness \cite{Filho3}. It is shown that this model exhibits a very rich phase diagram, including Berezinskii-Kosterlitz-Thouless (BKT) transitions, BKT endpoints, and isolated critical points. By means of Monte Carlo simulations, surface critical behaviors, including the corresponding critical exponents in the XY model, have been analysed in Ref. \cite{Landau2}.

In the present study, we investigate the metamagnetic XY model on a three-dimensional lattice by benefiting from Monte Carlo simulation with local spin update. As far as we know, the thermal and magnetic phase transition properties of the model mentioned above have not been studied yet. As mentioned before, the metamagnetic systems described by the Ising model in a magnetic field show us that they can include multi-critical points such as tri-critical points, bi-critical end points, and critical end points depending on the ratio between these ferromagnetic and antiferromagnetic interactions. Based on our numerical outputs, in a nutshell, we can say that the present system has very simple phase diagrams and exhibits antiferro-para transitions depending on the competition between different kinds of spin-spin interactions, and it does not include multi-critical points for the considered Hamiltonian parameters.

The paper is organized as follows: In the next section (section II) the model and Monte Carlo simulation scheme are discussed. The numerical results are reported in section III. The paper ends with a summary of the work in section IV.
\vskip 0.5 cm
\noindent {\large\bf II. The Model and Simulation method:}
\vskip 0.5 cm

\begin{figure}[h!]
\begin{center}
\includegraphics[angle=0,width=0.5\textwidth]{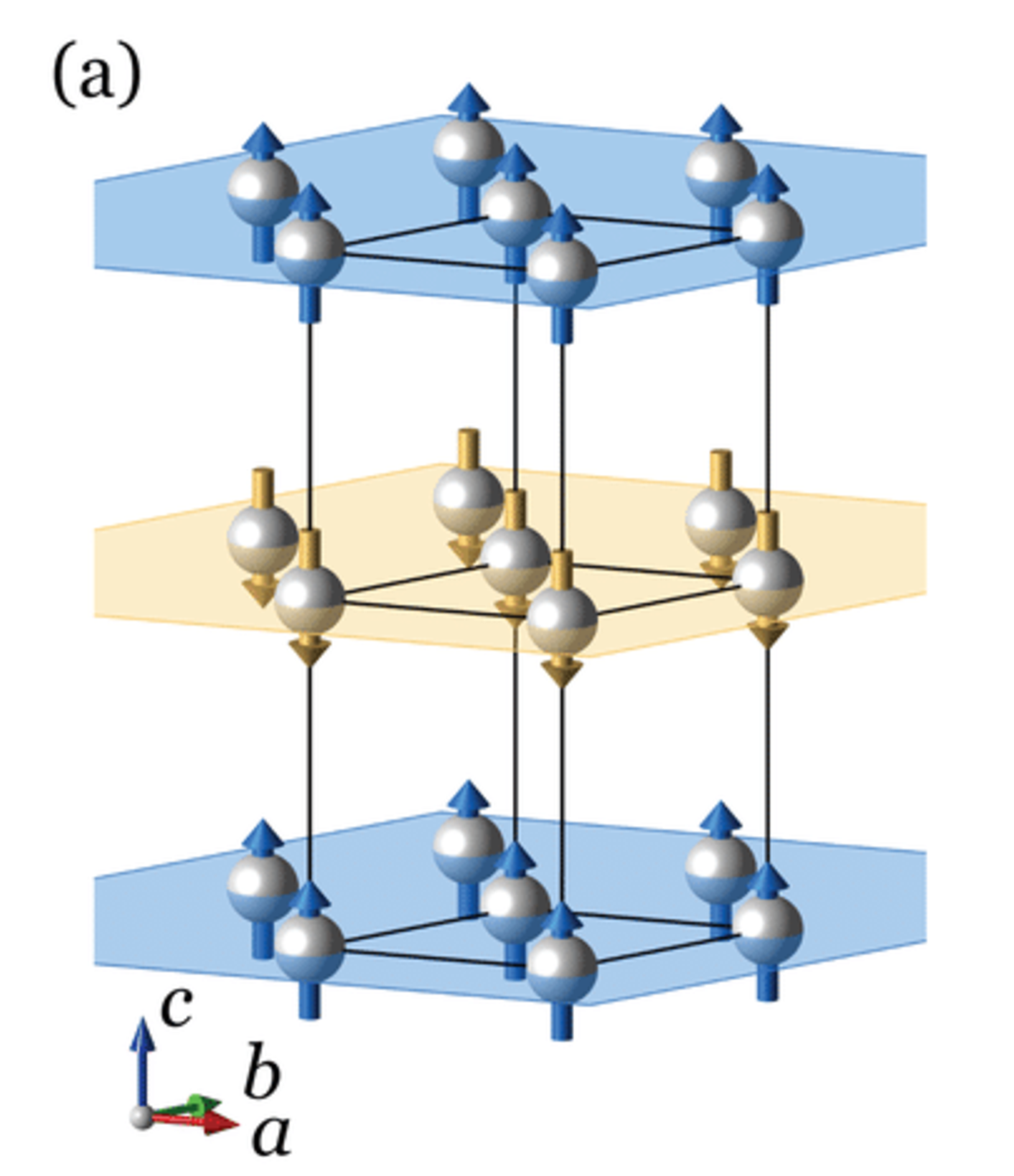}\\

\caption{A schematic of layered antiferromagnet.}
\label{image}
\end{center}
\end{figure}

\noindent
The spin-1 XY metamagnet can be modelled by the following Hamiltonian, \begin{equation} H= -{J_f}\sum_{intra-planar}(S_i^x S_j^x+S_i^yS_j^y) -{J_a}\sum_{inter-planar}(S_i^x S_j^x+S_i^yS_j^y) - \sum_{i}(h_xS_i^x+h_yS_i^y), \label{hamiltonian} \end{equation} where $S_i^x$ and $S_i^y$ are the $x$ and $y$-component of the spin ($|S|=1$) at $i$-th lattice site respectively. $J_f ( > 0)$ is the nearest neighbour intra planar ferromagnetic interaction strength and $J_a ( < 0)$ is the nearest neighbour inter planar antiferromagnetic interaction strength. Figure-\ref{image} is the schematic representation of such a layered antiferromagnetic ground state, where the spins are ferromagnetically ordered in a single layer. But the spins in adjuscent layers are antiferromagnetically ordered. Here the ratio ($R$) of the interaction strength is $R=-{{J_a} \over {J_f}}$. The $h_x$ and $h_y$ are the $x$ and $y$ components of the externally applied magnetic fields. The field is measured in the unit of $J_f$. Periodic boundary conditions are applied in all three directions.

In our simulation, we have considered $L \times L \times L$ simple cubic lattice of $L=40$. The initial state of the system was considered with high temperature random configurations of spins. This is paramagnetic phase corresponding to a high temperature ($T$) measured in the unit of $J_f/k_B$, where $k_B$ is Boltzmann constant. The system was updated by random updating scheme. In the random updating scheme, any lattice site (say $i$-th site) is chosen randomly where the old values of spin components ($S_i^x(old), S_i^y(old)$) have to be updated to another randomly chosen new values spin components ($S_i^x(new), S_i^y(new)$) according to the Metropolis probability\cite{simulation} \begin{equation} P(S_{old} \to S_{new}) = {\rm Min}\left[1, {\rm exp}\left({-\frac{\Delta H} {k_BT}}\right)\right], \label{metropolis} \end{equation} \noindent where $\Delta H$ (calculated from equation-\ref{hamiltonian}) is the change in energy due the change in spin configuration from old value to new value. If a random number (uniformly distributed between 0 and 1) is less than or equal to the Metropolis probability (equation-\ref{metropolis}), then the randomly chosen lattice site will be assigned the new values of spin components. In this way, $L^3$ such random updates are done. This constitutes a single Monte Carlo Step per Site (MCSS) and serves as the unit of time in this problem. Here, the last spin configuration for any temperature was considered as the initial configuration of the next lower temperature in the cooling process of the simulation.

Throughout the simulation we have considered $h_y=0$ and allowed $2\times10^5$ number of MCSS out of which we have discarded transient/initial $1.5\times10^5$ number of MCSS. The observable quantities are measured/calculated by averaging over $5\times10^4$ MCSS. We have checked that the initially discarded MCSS is sufficient to have equilibrium results within the accuracy of the interval of temperatures. In this way, just by cooling the system, from a high temperature paramagnetic phase, we have obtained various quantities as a function of temperature.

The present study is based on the proper characterization of various thermodynamic phases of the XY metamagnet. The different components of ferromagnetic and antiferromagnetic order parameters are responsible for representing the phases. The temperature at which the phase transition occurs can be determined by the thermal variations of the variance of different components of the order parameter (assumed to serve the role of susceptibility) and the specific heat. Keeping this in mind, we have calculated \cite{stanley} the following quantities:
\vskip 0.5cm

(1) X-component of the antiferromagnetic order parameter ($M_a^x$):

$m_a^x={{1} \over {L^3}}\left(\sum_{odd ~planes}S_i^x-\sum_{even ~planes}S_i^x\right)$

\begin{equation}
 M_a^x=\langle m_a^x\rangle = {1 \over \tau}\int m_a^x dt
\label{antimx}
\end{equation}

(2) Y- component of the antiferromagnetic order parameter ($M_a^y$):

$m_a^y=\displaystyle{{1} \over {L^3}}\displaystyle\left(\sum_{odd ~planes} S_i^y-\sum_{even ~planes}S_i^y\displaystyle\right)$

\begin{equation}
M_a^y=\langle m_a^y\rangle  =\displaystyle {1 \over \tau}\int m_a^y dt
\label{antimy}
\end{equation}

(3) X-component of ferromagnetic ordering ($M_f^x$):

$m_f^x=\displaystyle{{1} \over {L^3}}\left(\sum_{odd ~planes} S_i^x+\sum_{even ~planes}S_i^x\right)$

\begin{equation}
M_f^x=\langle m_f^x \rangle =\displaystyle {1 \over \tau}\int m_f^x dt
\label{ferromx}
\end{equation}

\noindent the symbol $\langle\cdots\rangle$ stands for time average over $\tau$ MCSS. The time average gives the result of ensemble 
average in the ergodic limit.

(4) Susceptibility of antiferromagnetic Y-ordering\cite{stanley}:

\begin{equation}
\chi_{ay}={{L^3 \left(\langle(m_a^y)^2\rangle-\langle(m_a^y)\rangle^2\right)} \over {k_B T}}
\label{chi}
\end{equation}

(5) Mean angle (odd plane)

\begin{equation}
\theta_o=\displaystyle{{2} \over {L^3}}\sum_{Odd ~planes}tan^{-1}\left({{S_i^y} \over {S_i^x}}\right)
\label{oang}
\end{equation}

(6) Mean angle (even plane)

\begin{equation}
\theta_e=\displaystyle {{2} \over {L^3}}\sum_{Even~ planes}tan^{-1}\left({{S_i^y} \over {S_i^x}}\right)
\label{eang}
\end{equation}

\newpage
\noindent {\large\bf III. Results:}
\vskip 0.5 cm

\vskip 0.5cm

\begin{figure}[h!]
\begin{center}
\includegraphics[angle=0,width=0.5\textwidth]{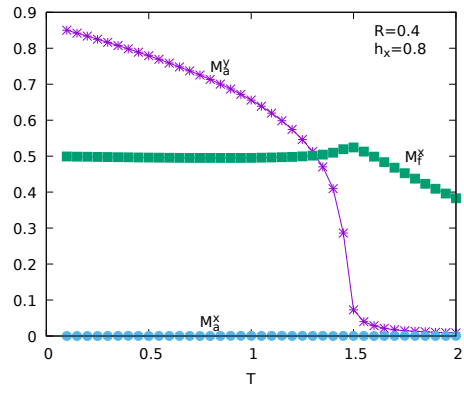}\\
(a)
\\
\includegraphics[angle=0,width=0.5\textwidth]{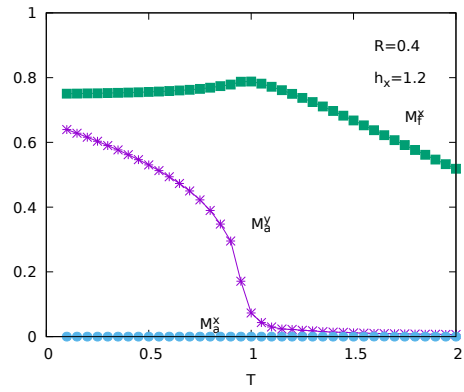}\\
(b)

\caption{The various ordering versus temperature.}
\label{order}
\end{center}
\end{figure}

The above mentioned (in the previsous section) thermodynamic quantities are studied as functions of temperature, as shown in the figures. Figure-\ref{order} represents the thermal variations of different order parameters. As the temperature of the system decreases (for $R=0.4$ and $h_x=0.8$), the antiferromagnetic ordering in the Y-direction $M_a^y$ (equation-\ref{antimy}) was found to take a nonzero value (Fig-\ref{order}(a)) at some transition temperature ($T_c$), whereas the antiferromagnetic ordering along the X-direction $M_a^x$ (equation-\ref{antimx}) remains zero for all temperatures. The ferromagnetic ordering in the X-directions $M_f^x$ (equation-\ref{ferromx}) assumes a nonzero value due to the application of an external magnetic field ($h_x$) in the X-direction. A significant value of the maximum of $M_f^x$ has been observed near the transition point. Since, the field $h_x$ is applied in the X-direction, the antiferromagnetic ordering in the X-direction is not possible. However, this favours the ferromagnetic ordering in the X-direction. This transition point was found to shift towards the low temperature for a larger value of the $h_x$ (Fig-\ref{order}(b)). This can be understood qualitatively as follows: the ferromagnetically (along the X-direction) ordering field $h_x$ favours $M_f^x$ to gets peaked near the antiferromagnetic (by $M_a^y$) transition temperature. The maximum of the ferromagnetic ($M_f^x$) ordering is accompanied by the antiferromagnetic ($M_a^y$) ordering here. This peak value of $M_f^x$ will be observed definitely at lower temperatures for higher values of $h_x$. So, the antiferromagnetic ordering (by $M_a^y$) will take place at lower temperatures for higher values of $h_x$.

\begin{figure}[h!]
\begin{center}
\includegraphics[angle=0,width=0.5\textwidth]{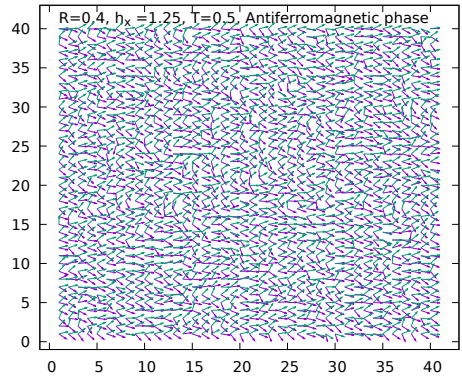}

\caption{The configurations of spins in the low temperature antiferromagnetic phase.The two colours represents two planes (8th and 9th plane).}
\label{spinmap}
\end{center}
\end{figure}

The configurations of the spins are studied for different sublattices (denoted by odd numbered planes and even numbered planes). A typical such spin configuration is shown in Fig-\ref{spinmap}. The spin configuration in the low temperature ordered phase shows that the values of $M_a^y$ is nonzero in the ordered phase. The value of $M_{a}^x$ remains zero for the entire range of the temperature, whereas the nonzero value of $M_f^x$ is also evident. Since the intra planar interaction is ferromagnetic and the inter planar interaction is considered antiferromagnetic in the model, it is quite natural that the spin configurations will be of the antiferromagnetic type in the adjacent planes. However, they will be oriented ferromagnetically in any single plane. A careful (with an enlarged picture) observation will display that the spins in different planes (depicted by two different colours (in Fig-\ref{spinmap}) are antiferromagnetically ordered (Y-component only) on an average. However, the spins in a single plane (any one colour here) are ferromagnetically ordered.

\begin{figure}[h!]
\begin{center}
\includegraphics[angle=0,width=0.5\textwidth]{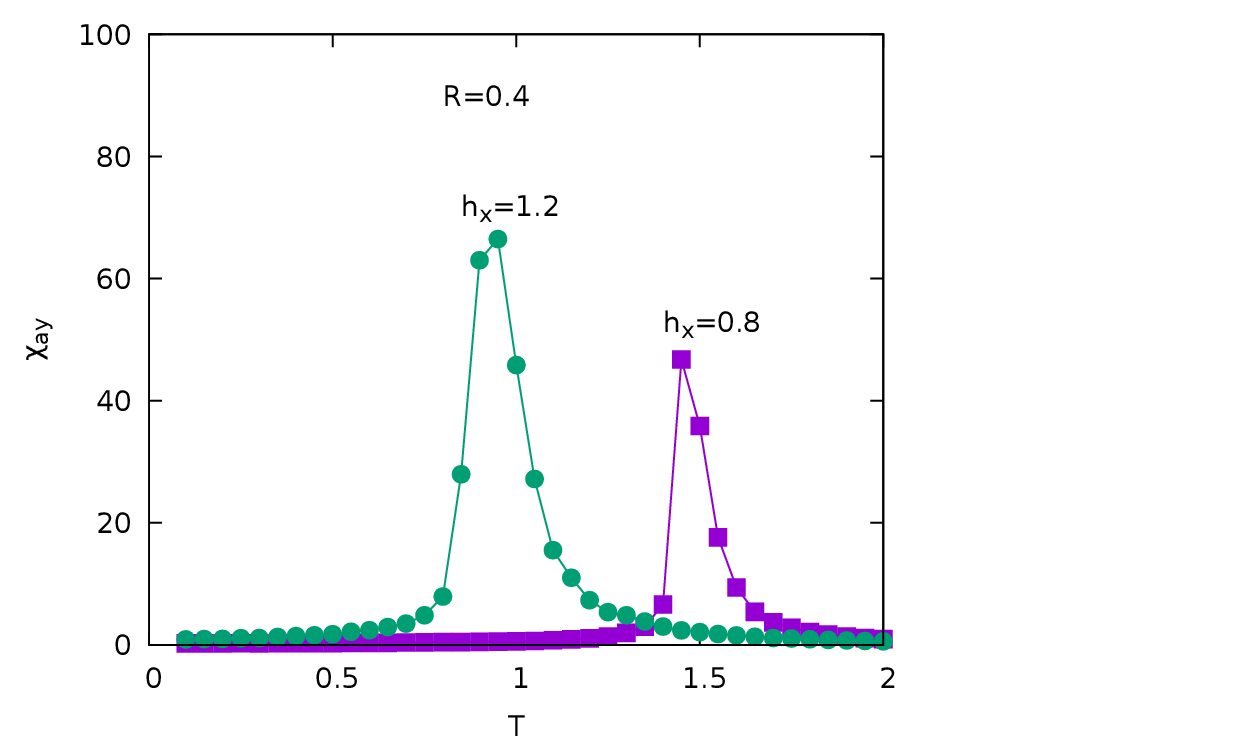}

\caption{The susceptibility $\chi_{ay}$ plotted against the temperature ($T$) for two different values of $h_x$.}
\label{susceptibility}
\end{center}
\end{figure}

The transition temperature for the different values of $h_x$ can also be determined from the temperature dependencies of the susceptibility $\chi_{ay}$ (the variance of $M_a^y$) of antiferromagnetic ordering. The $\chi_{ay}$ (calculated from the variances of $M_a^y$) are plotted against temperature for two different values of $h_x$ and shown in Fig-\ref{susceptibility}. The temperature which maximizes the susceptibility $\chi_{ay}$ is indicating the transition (from paramagnetic to antiferromagnetic ordering in Y-direction) temperature. This maximum of the susceptibility $\chi_{ay}$ is a finite size effect. In the thermodynamic limit ($L \to \infty$), this maximum is eventual divergence. This will be discussed later,  after a few paragraphs.

\begin{figure}[h!]
\begin{center}
\includegraphics[angle=0,width=0.5\textwidth]{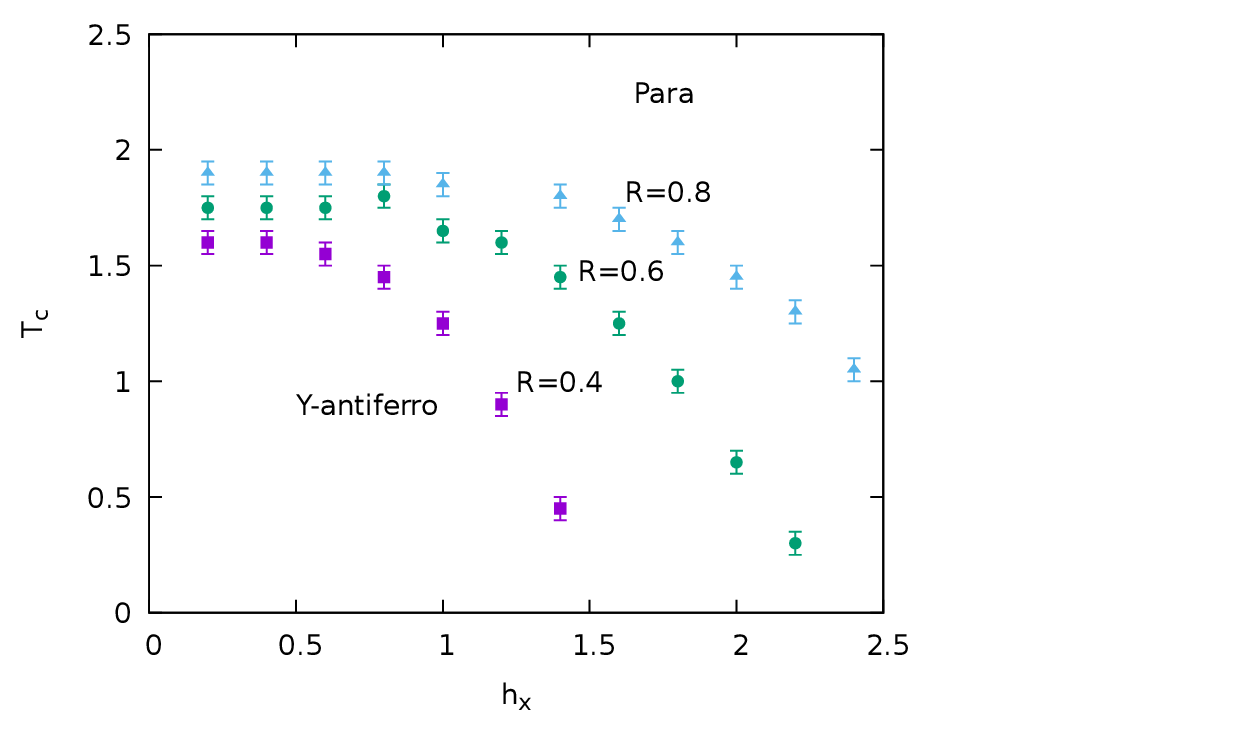}

\caption{The phase diagrams for different values of $R$.}
\label{phase}
\end{center}
\end{figure}

The results present that (for fixed $R$) the transition occurs at lower temperature for higher values of $h_x$ (see Fig-\ref{susceptibility}). The transition temperatures (for different values of $h_x$) are collected from the positions of peak of the susceptibility $\chi_{ay}$ (equation-\ref{chi}) studied as function of the temperature ($T$). In this way, the antiferromagnetic transition temperature can be found as a function of $h_x$. This (transition temperature) is fairly a monotonically decreasing function of $h_x$. The comprehensive phase diagram for the phase transition was drawn in the $h_x-T$ plane and shown in Fig-\ref{phase} for different values of $R$. Enclosed area of the region of ordered phase was found to increase as $R$ increased. This can be explained with physical argument as follows: for the stronger antiferromagnetic interaction (with fixed intra planar ferromagnetic interaction), the antiferromagnetic ordering will require more thermal fluctuations to become disordered. As a result, for fixed $h_x$, the antiferromagnetic transition will take place at higher temperatures for larger values of $R$. Hence, the phase boundary enclosed more regions of ordered phase for higher values of $R$. Before going further, at this point, we should note that we did not find any evidence of the metamagnetic anomalies \cite{Selke} for the three different values of $R$ parameters used in the study. The possible reason, for the absence of any metamagnetic anomaly, may be the continuous nature of the spin orientation in the XY model. The phase diagrams obtained here, are very simple,  and do not include any first-order phase transition points.

\begin{figure}[h!]
\begin{center}
\includegraphics[angle=0,width=0.5\textwidth]{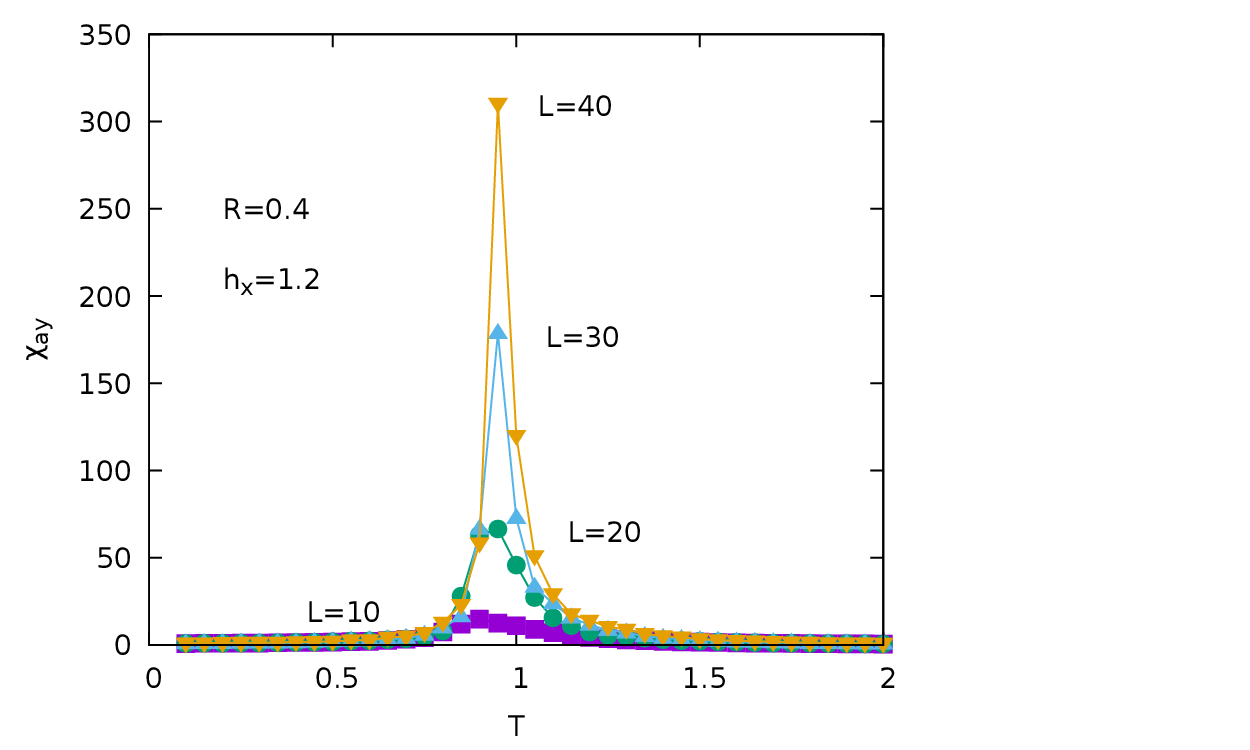}

\caption{The susceptibilities plotted against the temperature for
different $L$.}
\label{chi-T}
\end{center}
\end{figure}

Any equilibrium critical phenomenon shows the growth of correlations near the transition temperature. This is generally manifested in the growth of susceptibility near the transition point. The basic question is, whether this correlation provides long range ordering or not. To confirm the long range ordering via the critical growth of correlation, it is a standard technique to study the finite size analysis of the susceptibility. This has also been studied here. It is shown in Fig-\ref{chi-T}, where the susceptibility ($\chi_{ay}$) of the antiferromagnetic ordering ($M_a^y$) has been studied as function of temperature ($T$) for different system sizes ($L$). It is clear from the diagram that as the system becomes larger, the maximum value ($\chi^m_{ay}$) of the susceptibility ($\chi_{ay}$) increases. This confirms the critical growth of correlations and the existence of long-range order below the transition temperature.

\begin{figure}[h!]
\begin{center}
\includegraphics[angle=0,width=0.5\textwidth]{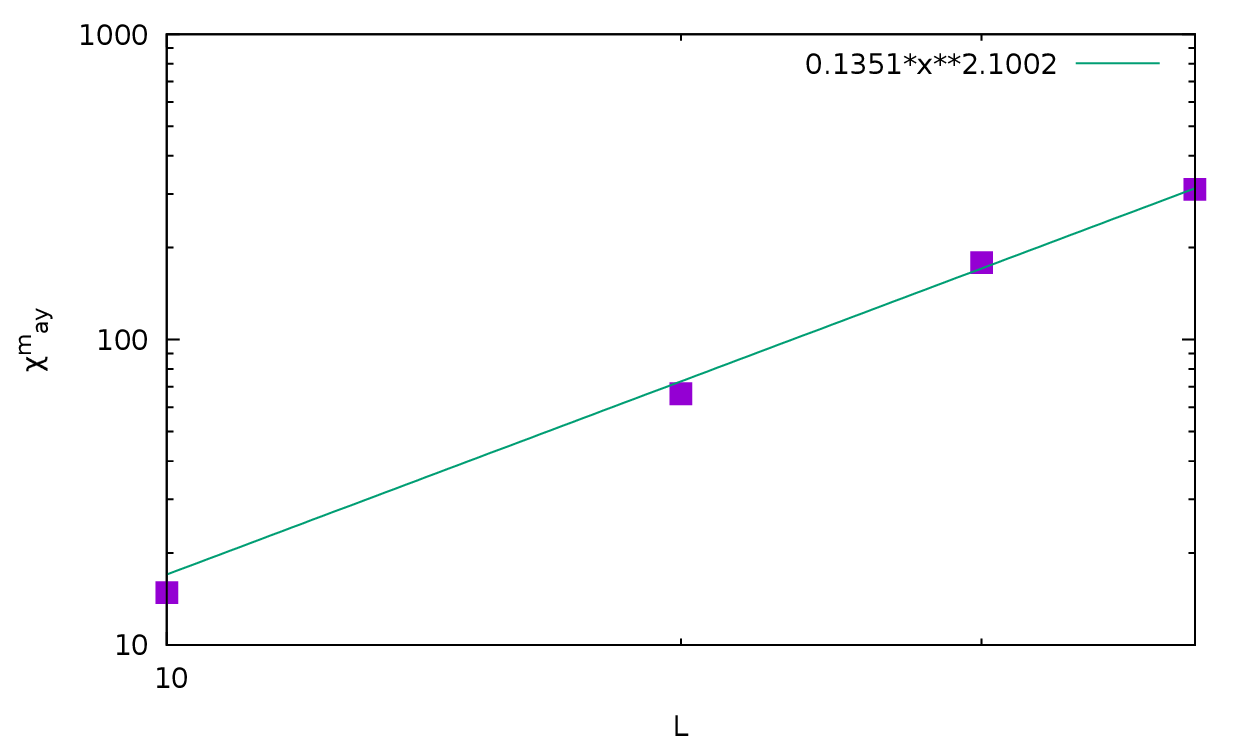}

\caption{The maximum values of the susceptibilities plotted against the
different sizes of the system $L$ in logarithmic scale.}
\label{maxchi-L}
\end{center}
\end{figure}

To be more specific in the mathematical form, we have performed the following study: Assuming the scaling law $\chi^m_{ay} \sim L^{({{\gamma} \over {\nu}})}$, the $\chi^m_{ay}$ has been plotted (in Fig-\ref{maxchi-L}) against $L$ (within our limited computational facilities) in logarithmic scales. It confirms the scaling behaviour with the estimated exponent ${{\gamma} \over {\nu}} = 2.10\pm0.11$. Furthermore, the value of the exponent (${{\gamma} \over {\nu}}$) is fairly close to that obtained\cite{univ} in the Monte Carlo calculations of the three dimensional (3D) XY ferromagnets. 

\begin{figure}[h!]
\begin{center}
\includegraphics[angle=0,width=0.5\textwidth]{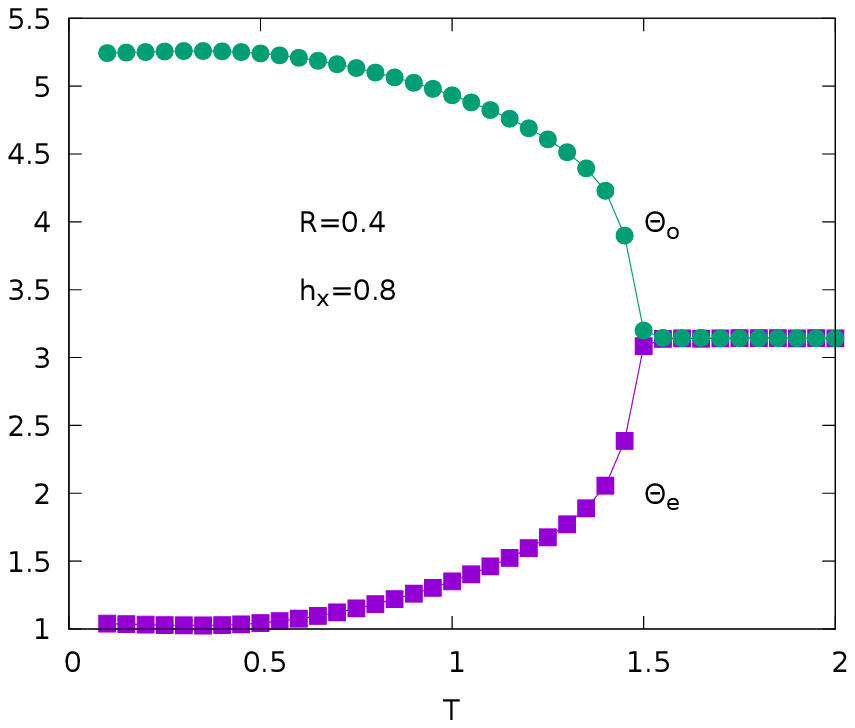}\\
(a)
\\
\includegraphics[angle=0,width=0.5\textwidth]{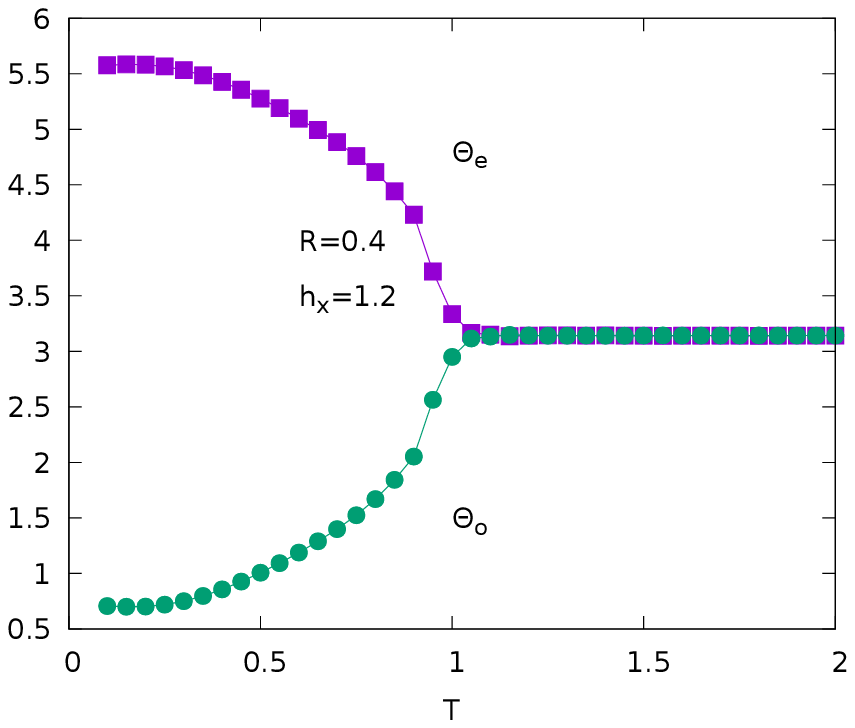}\\
(b)
\\
\includegraphics[angle=0,width=0.5\textwidth]{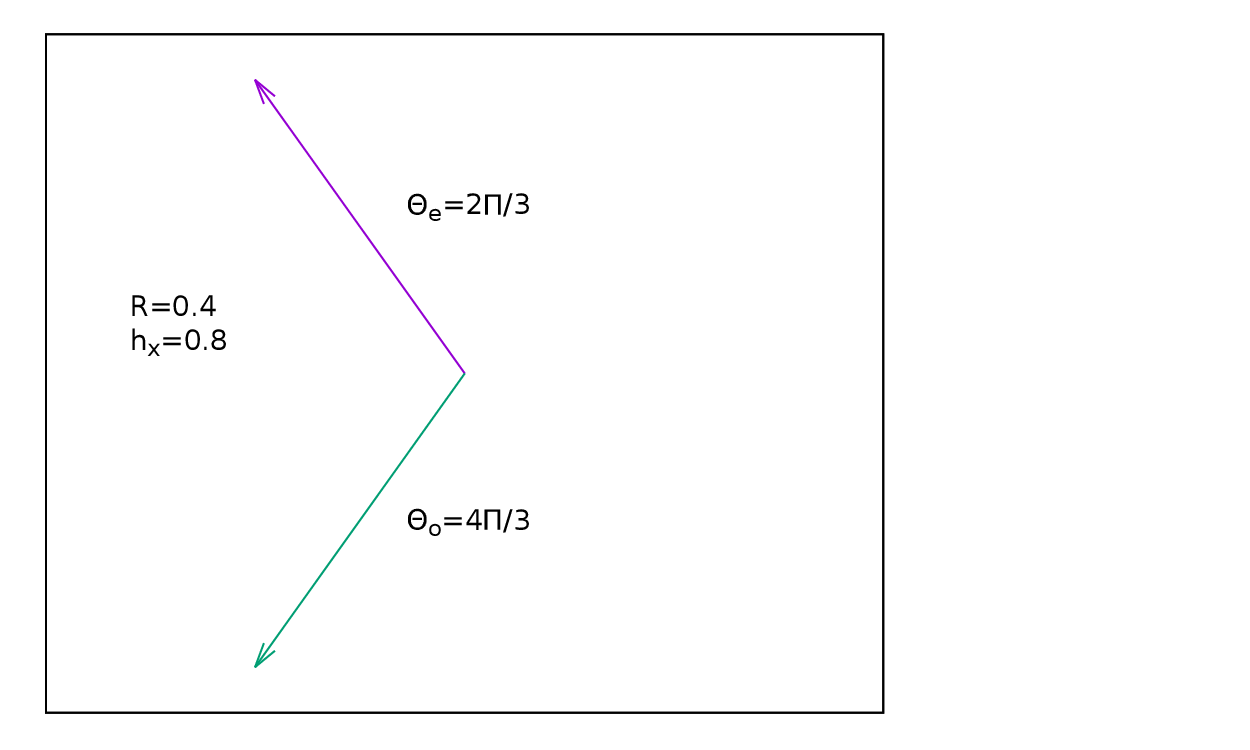}\\
(c)
\caption{The average angles of odd ($\theta_o$) and even ($\theta_e$) planes for R=0.4 for two different values of $h_x$. (a) For $h_x$=0.8 and (b) $h_x=1.2$. (c) shows
a demonstration of average angles for even and odd planes. }
\label{angles}
\end{center}
\end{figure}

Can one get any alternative way to find the antiferromagnetic phase transition temperature in the model considered here ? For SO(2) symmetric spin models (XY-model), it can also be found by studying the mean angles of two sublattices, namely, $\theta_o$ (equation-\ref{oang}) for odd numbered planes and $\theta_e$ (equation-\ref{eang}) for even numbered planes. The angle of a spin vector is measured with respect to the positive X-axis. In the paramagnetic phase, the angles are random and uniformly distributed between $0$ and $2\pi$. So, the mean sublattice angles (for two different sublattices) are the same and approximately equal to $\pi$. But below the critical temperature, the mean angles of different sublattices get widely apart, confirming the antiferromagnetic ordering in the Y-direction. These are shown in Fig-\ref{angles}. The deviations of the values of mean angle from $\pi$ is the signature of antiferromagnetic ordering in this model. Fig-\ref{angles}(c) demonstrates a typical case of average angles of even and odd planes. For $R=0.4$, $h_x=0.8$ and close to $T=1.35$ (Fig-\ref{angles}(a)), the mean angle of the spins in even planes is $2\pi/3$ and that for the odd planes is $4\pi/3$. This configuration confirms the antiferromagnetic ordering in Y-direction ($M_a^y \neq 0$). It may be noted from Fig-\ref{angles}(a) and Fig-\ref{angles}(b) that the antiferromagnetic transition may also be detected from the study of the mean angles. The deviation of the mean angles, from that of $\pi$ , indicates the antiferromagnetic transition. Here also, from the Fig-\ref{angles}, it is observed that the transition occurs at lower temperatures for higher values of $h_x$.
\vskip 3 cm
\noindent {\large\bf IV. Summary:}

\vskip 0.2 cm

Let us summarise our main findings from the Monte Carlo study of the layered antiferromagnetic XY model in the presence of an external magnetic field applied in the x-directions.

Depending on the value of $h_x$ and $R$ the system undergoes equilibrium phase transitions. The antiferromagnetic ordering, along the Y-direction,  grows at a critical temperature. This critical temperature is obtained from the position of the peak of susceptibility, $\chi_{ay}$, plotted as a function of temperature ($T$).

The comprehensive phase boundary, of the antiferro-para transition, shows a significant dependence on the different values of the ratio of relative interaction strength ($R$). The ordered region, bounded by the phase boundary, was found to expand as the ratio of the relative interaction strength ($R$) increased. The growth of the correlation was indeed observed near the transition temperature. The height of the susceptibility has been found to increase as the size of the system increases. Assuming the scaling law $\chi^m_{ay} \sim L^{({{\gamma} \over {\nu}})}$, the exponent ${{\gamma} \over {\nu}}=2.10\pm0.11$ has been estimated for $h_x = 1.2$ and $R = 0.4$, in our present simulational study. This value is very close to that mentioned in Ref\cite{univ} for the 3D XY universality class. The only reason for choosing such values of $h_x$ and $R$, was to estimate the expected critical exponent and universality class of the considered model. A similar analysis can be performed for all phase transition points appearing in the phase diagrams. However, it is known that the universality classes are not dependent on the Hamiltonian parameters\cite{erol}. The critical exponents to be obtained for other sets of Hamiltonian parameters may change (numerically) a bit, but we believe that this would not change the expected universality class of the present model within errors. Further and dedicated simulations are needed to obtain the critical exponents with high precision, but they are beyond the scope of the current study.

\vskip 2 cm
\noindent {\large\bf V. Acknowledgements:}
	
MA would like to acknowledge the FRPDF research grant provided by Presidency University. Some of the of the numerical calculations reported in this paper were performed at TÜBITAK ULAKBIM (Turkish agency), High Performance and Grid Computing Center (TRUBA Resources).

\newpage

\newpage

\newpage

\newpage

\newpage

\newpage

\newpage

\newpage


\begin{thebibliography}{<num>}
\bibitem{Moreira1} A.F.S. Moreira, W. Figueiredo and V.B. Henriques, Eur. Phys. J. B \textbf{27}, 153 (2002).
\bibitem{Moreira2} A.F.S. Moreira, W. Figueiredo and V.B. Henriques, Phys. Rev. B \textbf{66}, 224425 (2002).
\bibitem{Liu} J. Liu, G. Wei and H. Miao, J. Magn. Magn. Mater. \textbf{315}, 101 (2007).
\bibitem{Wei} G. Wei, J. Liu, H. Miao and A. Du, Phys. Rev. B \textbf{76}, 054402 (2007).
\bibitem{Liang} Y.-Q. Liang, G.-Z. Wei, X.-J. Xu and G.-L. Song, J. Magn. Magn. Mater. \textbf{322}, 2219 (2010).
\bibitem{Gulpinar1} G. Gulpinar and Y. Karaaslan, Phys. Lett. A \textbf{375}, 978 (2011).
\bibitem{Gulpinar2} G. Gulpinar and E. Vatansever, J. Magn. Magn. Mater. \textbf{324}, 3784 (2012).
\bibitem{Nascimento} D.A. do Nascimento, J.T. Pacobahyba, M.A. Neto, O.R. Salmon and J.A. Plascak,
Physica A \textbf{474}, 224 (2017).



\bibitem{Zukovic0} M. \v{Z}ukovic and A. Bob\'{a}k, J. Magn. Magn. Mater. \textbf{170}, 49 (1997).
\bibitem{Zukovic01} M. \v{Z}ukovic, A. Bob\'{a}k and T. Idogaki, J. Magn. Magn. Mater. \textbf{188}, 52 (1998).
\bibitem{Filho} E.B. Filho and J.D. de Sousa, Phys. Lett. A \textbf{323}, 9 (2004).
\bibitem{Geng} J. Geng, G. Wei and H. Miao, J. Magn. Magn. Mater. \textbf{320}, 1010 (2008).
\bibitem{Miao} H. Miao, G. Wei, J. Liu and J. Geng, J. Magn. Magn. Mater. \textbf{320}, 2172 (2008).


\bibitem{Landau} D.P. Landau, Phys. Rev. Lett. \textbf{28}, 449 (1972).
\bibitem{Aora} B.L. Aora and D.P. Landau, AIP Conf. Proc. \textbf{10}, 870 (1973).
\bibitem{Hernandez} L. Hern\'{a}ndez, H.T. Diep and D. Bertrand, Phys. Rev. B \textbf{47}, 2602 (1993).
\bibitem{Pleimling1} M. Pleimling and W. Selke, Phys. Rev. B \textbf{56}, 8855 (1997).
\bibitem{Pleimling2} M. Pleimling and W. Selke, Phys. Rev. B \textbf{59}, 8395 (1999).
\bibitem{Pleimling3} M. Pleimling, Eur. Phys. J. B \textbf{10}, 465 (1999).
\bibitem{Acharyya1}  M. Acharyya, U. Nowak and K.D. Usadel, Phys. Rev B \textbf{61}, 464 (2000).
\bibitem{Zukovic1} M. \v{Z}ukovic and T. Idogaki, Phys. Rev. B \textbf{61}, 50 (2000).
\bibitem{Zukovic2} M. \v{Z}ukovic and T. Idogaki, J. Magn. Magn. Mater. \textbf{208}, 120 (2000).
\bibitem{Weinzenmann} A. Weinzenmann, M. Godoy and A.S. de Arruda, Braz. J. Phys. \textbf{36}, 645 (2006).
\bibitem{Chou} Y.-L. Chou and M. Pleimling, Phys. Rev. B \textbf{84}, 134422 (2011).
\bibitem{Mayberry} J. Mayberry, K. Taucher and M. Pleimling, Phys. Rev. B \textbf{90}, 014438 (2014).

\bibitem{Harbus1} F. Harbus and H.E. Stanley, Phys. Rev. B \textbf{8}, 1141 (1973).
\bibitem{Harbus2} F. Harbus and H.E. Stanley, Phys. Rev. B \textbf{8}, 1156 (1973).
\bibitem{Stryjewski} E. Stryjewski and N. Giardano, Advances in Physics \textbf{26}, 487 (1977).
\bibitem{Keskin1} M. Keskin, O. Canko and E. Kantar, Phys. Lett. A \textbf{373}, 2201 (2009).
\bibitem{Deviren1} B. Deviren, M. Keskin, Phys. Lett. A \textbf{374}, 3119 (2010).
\bibitem{Temizer} \"{U}. Temizer, Chin. Phys. B \textbf{23}, 070511 (2014).
\bibitem{Acharyya2} M. Acharyya, J. Magn. Magn. Mater. \textbf{382}, 206 (2015).
\bibitem{Kosterlitz} J.M. Kosterlitz and D.J. Thouless, J. Phys. C: Solid State Phys. \textbf{6}, 1181 (1973).
\bibitem{Krech} M. Krech and D.P. Landau, Phys. Rev. B \textbf{60}, 3375 (1999).
\bibitem{Filho2} J.B. S.-Filho and J.A. Plascak, Comput. Phys. Commun \textbf{182}, 1130 (2011).
\bibitem{Filho3} J.B. S.-Filho and J.A. Plascak, Phys. Rev. E \textbf{96}, 032141 (2017).
\bibitem{Landau2} D.P. Landau, R. Pandey and K. Binder, Phys. Rev. B \textbf{39}, 12302 (1989).
\bibitem{simulation} D.P. Landau and K. Binder, A Guide to Monte Carlo
Simulations in Statistical Physics (Cambridge University
Press, Cambridge, England, 2000).
\bibitem{stanley} H. E. Stanley, {\it Introduction to phase transition and
critical phenomena}, Clarendon Press, Oxford, 1971.
\bibitem{Selke} W. Selke, Z. Phys. B \textbf{101}, 145 (1996).
\bibitem{univ} M. Campostrini, M. Hasenbusch, A. Pelissetto, P. rossi and E. Vicari, Phys. Rev. B, {\bf 63}, 214503 (2001).
\bibitem{erol} E. Vatansever and N. G. Fytas, Phys. Rev. E, {\bf 97}, 012122 (2018).

\end{thebibliography}
\end{document}